# Challenges and perspectives for new material solutions in batteries

Vittorio Pellegrini[a,b], Silvia Bodoardo[c], Daniel Brandell[d], Kristina Edström[d]

[a] IIT Graphene Labs, Istituto Italiano di Tecnologia, Via Morego 30, 16153, Genova, Italy
[b] Bedimensional S.p.a. Via Albisola 121, 16153, Genova, Italy
[c] Electrochemistry group, Department of Applied Science and Technology, Politecnico di Torino, Torino, Italy
[d] Department of Chemistry – Ångström Laboratory, Uppsala University, Box 538, SE-751 21, Uppsala, Sweden

**We outline main challenges for future research in batteries, particularly, addressing the urgent needs of developing new environmentally friendly material solutions to enhance the energy density and safety of these storage devices. This will require embracing a multidisciplinary approach encompassing traditional electrochemistry and experimental solid-state physics, multiscale computational modelling, materials synthesis, and advanced characterization and testing**

Alessandro Volta was the first in 1796 to demonstrate a practical battery, the so called Voltaic pile, capable to exploit the energy delivered by spontaneous chemical redox reactions to produce electric power in a controlled fashion[1,2]. Batteries have evolved significantly since Volta's time, and today the use of rechargeable Li-ion batteries is pivotal in a wide range of fields, such as electric or hybrid vehicles, portable devices, smart energy grids and stationary devices for renewable energy[3]. The battery is an electrochemical cell composed by three crucial elements: a negative electrode (anode) able to accommodate ions during charging and to release electrons to the external circuit during discharge; a positive electrode (cathode) which is reduced during discharge; and an electrolyte solution containing dissociated salts, which enable ion transfer between the two electrodes. Each of these components are vital to develop to perfection in order to reach the high performances requested by consumers. Today, a commercial Li-ion battery consists of a graphite-based anode[4], a mixture of carbonate solvents containing a lithium salt as electrolyte[5] and a cathode comprising active materials, such as $LiCoO_2$ which is extensively used in portable devices, or the NMC ($LiNi_xCo_yMn_zO_2$; where x,y,z can vary in relation to each other with the total sum of one), which is employed in electric vehicles. These batteries are able to work with a voltage of the order of 3.5-4.0 V reaching a specific energy of around 250 Wh kg$^{-1}$ while keeping an overall efficiency above 80% for hundreds or thousands of cycles. Nevertheless, the increasing demands for applications requiring superior energy storage performances, such as in the automotive market which is still dominated by internal combustion engines based on fossil fuels (1 liter diesel corresponds to $10^4$ Wh), have originated in an increasing interest in developing alternative material solutions for Li-ion batteries. An additional target is to increase the safety and the sustainable aspects of the battery, which motivates the development of, for instance, solid-state electrolytes. Different energy storage systems, based on low-cost cathodes such as sulfur in the lithium-sulfur battery, with a theoretical specific energy density in a full cell of about 2500 Wh kg$^{-1}$ (moreover, elemental sulfur is abundant in nature, cheap and sustainable)[6] or metal-air batteries with a theoretical energy density of about 11000 Wh kg$^{-1}$ (referring to the Lithium anode) in the case of Li-air batteries,[7] have also emerged as intense research topics. Both these battery chemistries are highly challenging and the practical capacities obtained are far from the theoretical values, which makes scientific efforts important and timely. Figure 1 reports a bibliometric analysis aimed to show how the interest in different types of battery chemistries and methods has grown in academic science from the seminal paper on intercalation processes in Li-ion batteries of Whittingham in 1976[8].

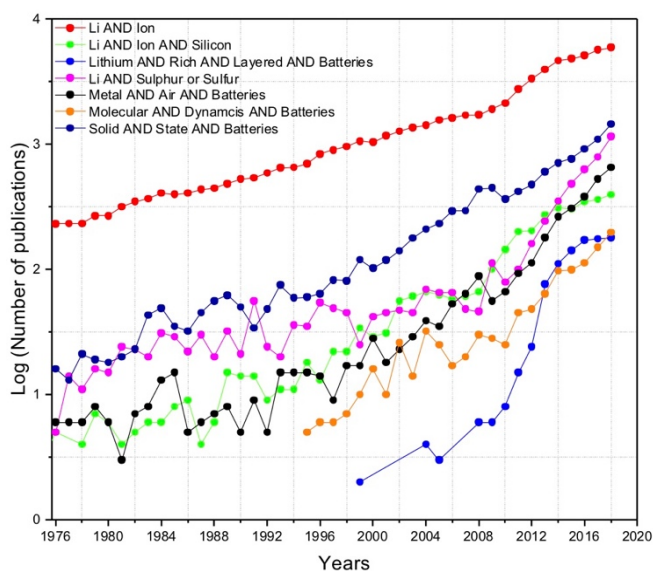

Figure 1. Number of publications as a function of year in the period 1976-2018 following SCOPUS. Different curves correspond to searches with words in title or abstract or keywords as reported in the inset.

Concerning Li-ion batteries, the anodic material represents a first challenge to address. Graphite is able to exchange only 1 Li$^+$ per every 6 C atoms, through an intercalation-deintercalation mechanism, and therefore it offers a limited specific capacity (expressed in delivered current (A) per hour (h) per gram (g) of the active materials) up to theoretically 372 mAh/g and practically around 350 mAh/g. For this reason, alternative materials that exploit other processes such as alloying[9] or conversion[10] are currently extensively studied[11]. Silicon, in particular, represents one of the most promising alloying anode materials to replace graphite-based anodes[12,13,14]. Silicon can theoretically react in a lithium cell through an alloying reaction to form Li$_{4.4}$Si, achieving a maximum specific capacity of 4200 mAhg$^{-1}$. However, several issues affect silicon, as well as other alloying materials, hindering its implementation in commercial batteries. One of the main issues of silicon is associated to the significant volume changes (>300%) during the lithiation/delithiation processes[15]. The volume change produces cracks and pulverization of the electrode, leading to loss of electrical contact and consequently poor cycling performance. Another issue is the reactions with the surface of the silicon particle and the electrolyte where the surface is not passivated by a stable solid electrolyte interphase (SEI). In order to overcome the aforementioned issues, several strategies can be adopted mainly addressing the design of advanced electrode structures[16], including, in particular, the exploitation of silicon nanoparticles[17] and their encapsulation into a carbon matrix of for example graphene to prevent direct contact with the electrolyte[18,19,20,21,22]. Alternatively, thin layers of an oxide protecting the surface can be employed, for example, TiO$_{2-x}$F$_x$,[23] or the electrode binder system can be tailored to prolong battery cycling[24]. Other two-dimensional materials than graphene, such as for example exfoliated black phosphorous[25] that in its three-dimensional structure has a theoretical capacity of 2596 mAhg$^{-1}$,[26] and the transition metal carbides and nitrides (MXenes), represent emerging material solutions for batteries[27,28] in combination with silicon[29] and other active materials.

It is worth mentioning that Li metal, which is an ideal candidate for anodes thanks to its high theoretical specific capacity (3,860 mAh g$^{-1}$) and very low redox potential, has re-emerged as a practical solution stimulated by the development of high-capacity cathodic materials. Ways to stabilize Li metal anodes and the development of new strategies for protecting the Li metal for long-term stable cycling are current topics of research [30,31].

An additional challenging topic for Li-ion batteries is that of replacing the cathode active materials, such as the already mentioned LiCoO$_2$ and NMC. These layered oxides have been extensively studied in terms of electronic and structural properties[32]. However, they are approaching their intrinsic limit in terms of energy density with respect to the cathode mass; i.e., 550 Wh kg$^{-1}$ and 800 Wh kg$^{-1}$ respectively[33,34]. Moreover, the use of expensive metals and the toxicity of cobalt compounds, which are required in the production processes has in the last years triggered increasing interest in identifying new cathode materials (see blue dots in Fig.1) capable to fulfill the high demand of energy density and environmental compatibility[35]. To this end, a new class of Li-rich layered oxides having a generic formula Li$_{1+x}$M$_{1-x}$O$_2$, in which M is a mix of transition metals as Ni, Mn, Co, and Co is simultaneously (partially) replaced by Mn, Ni and Li [36,37,38,39] are emerging as promising solutions that can deliver high specific capacities and/or discharge voltages of >4.5 V[40]. Their practical use in a Li-ion device is still limited by several issues, including large irreversible capacity losses in the first cycles, poor rate capability and gradual voltage decay during cycling. Some of these problems are associated with structural changes from layered structures to spinel-like structures[41] during extended cycling. Future improvements might require the exploration of different synthetic strategies, including doping with other metal (Al, Zr, Fe, Cr, Mg) and surface modifications and coatings. Another class of promising cathode crystals are the manganese oxide-based spinel structures, such as, for example, the Ni-substituted material LiNi$_{0.5}$Mn$_{1.5}$O$_4$ because of its high operational voltage (up to ~4.7 V vs. Li$^+$/Li$^0$) and fast three-dimensional Li$^+$ diffusion channels.

Another challenge in future battery research is that of replacing the liquid battery electrolyte with solid-state counterparts leading to a solid-state battery. In this way, two major advantages become possible: firstly, solid electrolytes are significantly less flammable which contributes to a much safer battery; secondly, the solid electrolyte can be made very thin, which renders a higher energy density[42,43]. This motivates a continuously-growing interest in solid-state batteries (see dark blue dots in Fig.1). A third advantage, not frequently highlighted in literature, is that solid electrolytes allow tailoring specific electrolyte systems at the anode and cathode, respectively[44]. The two different categories of solid electrolytes which dominate the field – ceramic and solid polymer electrolytes – struggle with different problems. Ceramic electrolytes, primarily phosphosulfides, garnet oxides, perovskites and different types of glasses such as lithium phosphorus oxynitride (LiPON)[45,46] not seldom possess reasonably high ionic conductivities, but can at the same time be brittle, and their rigid structures often yields high interfacial resistance. Lithium battery electrodes are often porous, and it becomes difficult for a solid ceramic to ensure good surface contact with every surface point of the active material throughout the electrode. Polymer electrolytes, traditionally based on polyethers[47] and more recently on polyesters[48] and ionomers[49], are in this perspective advantageous and can accommodate the active material's volume changes during battery cycling, but on the other hand struggle with lower ionic conductivities, especially at ambient temperatures. That the problems for ceramic and polymer electrolytes do not overlap to a vast extent means that synergistic effects are highly possible. It is therefore not far-fetched to envision next-generation electrolytes as solid composites of these two types of materials[50].

Another strategy to surpass the intrinsic limitations of any cathodic materials in Li-ion batteries is represented by the Li-sulfur or the metal-air technologies. Concerning Li-S batteries, the spontaneous sulfur reaction with lithium passes through several reaction steps with the production of radicals and compounds which in the total reaction formula can be expressed as $S_8 + 16Li^+ + 16\ e^- \rightarrow 8\ Li_2S$. The pristine octa-sulfur ring reacts with lithium ions during the discharge processes, producing lithium-polysulfides with different chain lengths and finally generating solid $Li_2S$ species at the end of discharge[51]. This conversion reaction is highly complex due to the phase changes of the sulfuric species, from solid to liquid, during the voltage drop in the discharge profile from 2.4 V (solid region) to 2.1 V, when they are mostly liquid[52]. The sulfur battery shows outstanding performances with respect to commercial Li-ion battery cathodes in terms of capacity, but many issues remains before large-scale implementation, not least the low electronic conductivity in the pristine state and the poor electron charge transfer in the liquid phase[53]. Moreover, the polysulfides are soluble in common electrolytes, thereby leading to loss of active sulfur material, decrease of the specific capacity and possible cell failure[54]. Furthermore, the polysulfides may migrate from the cathode to the anode side and then react on the anode surface, producing an electrochemical short-circuit known as the polysulfide shuttle effect, which effectively self-discharges the battery[55,56]. The most common solution, in order to overcome the low conductivity and the disintegration of the lithium polysulfide, is the use of a carbonaceous matrix for the sulfur that allows high electron charge transfer trough the electrode and acts as a substrate in order to retain the polysulfide in the cathodic region of the cell[57,58]. Many carbonaceous structures have been proposed to solve these issues by the preparation of sulfur-carbon composites: carbon nanotubes[59], mesocarbon microbeads (MCMB carbon)[60], carbon nanosheets[61], amorphous carbon[62], graphite[63], or graphene[64,65], improving the performances and the applicability of the lithium sulfur battery in terms of better stability, longer cycle life and higher energy density[66].

However, most of the preparation pathways of the cathodes are complex and involve expensive procedures. This, together with the need to keep at the minimum weight% value, at most 30%, the amount of carbon in the electrode, [67]is stimulating further growing research on materials and material production combinations [68] (see pink dots in Fig.1) in order to overcome these major obstacles towards the spread of the Li-S storage technology.

For Li-air batteries, several issues have to be overcome, mainly related to the strong reactivity of the reaction products formed when lithium reactions with oxygen[69], for example $Li_2O_2$, which decompose the electrolyte and generate clogging of the cathode pores[70]. The most promising positive electrode is a carbon-based matrix which can contain catalysts (e.g., Pd, $MnO_x$, etc.) to reduce problems related to the sluggish oxygen redox kinetics[71,72]. An efficient oxygen cathode should satisfy a number of requirements, such as electrical conductivity, appropriate porosity to store the solid discharge products, promotion of ORR / OER processes with bi-functional catalysts while maintaining chemical stability towards radicals and nucleophilic intermediates and, finally, satisfactorily wettability by the electrolyte[73]. In particular, to meet the last two requirements, it is important to consider the role of the binders used in cathode preparation since they also are unstable to the superoxide ions formed during reduction[74,75]. For these reasons, self-standing or binder-free cathodes have been proposed for $Li-O_2$ cells[76]. In order to use air with relative humidity, it is necessary to develop selective membranes allowing longer cyclability[77]. Moreover, it is worth to mention that sodium–oxygen ($Na-O_2$) batteries are arising as a promising alternative due to key advantages of Na such as low cost, high abundance and better ionic conductivities (relative to Li)[78,79], although with somewhat lower specific energies than in $Li-O_2$ batteries. Reduced graphene oxide aerogels have been proposed as promising self-standing, binder-free cathodes for such $Na-O_2$ batteries[80]. Figure 2 summarizes the different material combinations that can be exploited for the next generation of Lithium-based batteries.

Finally, in the search of new material solutions and in the development of next-generation smarter/safer batteries, we can expect that an increasing role will be played by modern multiscale computation approaches (see orange points in Fig.1). In combination with artificial intelligence/machine learning mechanisms,[81,82] and possibly with the aid of quantum computing,[83] the materials research community will be capable of performing new autonomous discoveries and interphase engineering, and thereby to guide research and development activities through linking the fundamental and the applied levels.

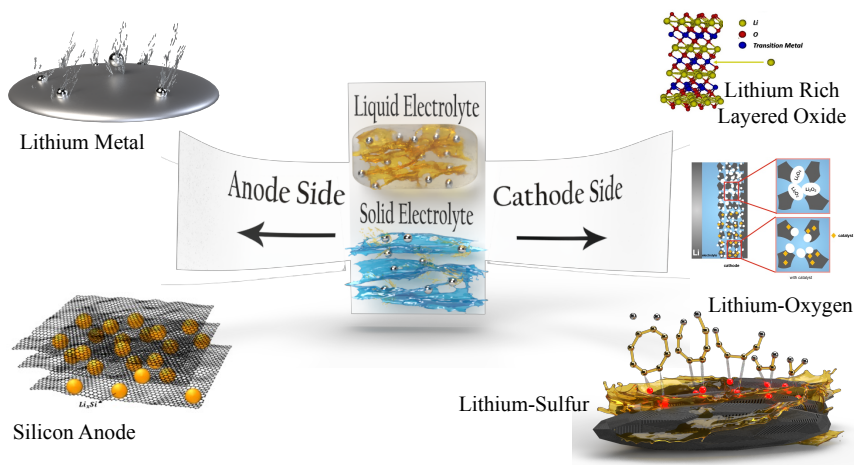

Figure 2. Illustration of the different materials for Lithium-based batteries as discussed in this paper. Silicon-carbon-graphene and Lithium metal as anodes, carbon-sulphur composites, Li-rich layered oxides and carbon-based matrix which can contain catalysts as cathodes. These materials can be combined with both liquid and solid electrolytes to form Li-ion, Li-Sulphur and Li-air batteries.

Acknowledgements: We thank Lorenzo Carbone and Arcangelo Celeste for their support during the writing of this manuscript. VP would like to thank the support of the European Union's Horizon 2020 research and innovation program under grant agreement no. 785219, GrapheneCore2. KE and DB thank the Swedish Strategic Research Area StandUp for Energy.

[1] A. Volta, *Philos Trans* **2** 402 (1800)
[2] B. Scrosati, *Journal of Solid State Electrochemistry* **15**, 1623 (2011)
[3] E.M. Erickson, C. Ghanty, and D. Aurbach, *J. Phys. Chem. Lett.* **5**, 3313 (2014)
[4] D. Aurbach, E. Zinigrad, Y. Cohen, and H. Teller, *Solid State Ionics* **148**, 405 (2002)
[5] M. Dahbi, F. Ghamouss, F. Tran-Van, D. Lemordant, and M. Anouti, *J. Power Sources* **196**, 9743 (2011)
[6] A. Manthiram, Y. Fu, S. Chung, C. Zu, and Y. & Su, *Chem. Rev.* **114**, 11751 (2014)
[7] P. Tan, H.R. Jiang, X.B. Zhu, L. An, C.Y. Jung, M.C. Wu, L. Shi, W. Shyy, and T.S. Zhao *Applied Energy* **204** 780 (2017)
[8] S. Whittingham, *Science* **192**, 1126 (1976).
[9] M.N. Obrovac, and V.L. Chevrier, *Chem. Rev.* **114**, 11444 (2014)
[10] P. Poizot, S. Laruelle, S. Grugeon, L. Dupont, J.M. Tarascon, *Nature* **407**, 496 (2000)
[11] J.W. Choi, D. Aurbach, *Nat. Rev. Mat.* **1**, 16013 (2016)
[12] M.N. Obrovac, and V.L. Chevrier, *Chem. Rev.* **114**, 11444 (2014)
[13] A. Casimir, H. Zhang, O. Ogoke, J.L. Amine, J. Lu, and G. Wu, *Nano Energy* **27**, 359 (2016)
[14] B. Liang, Y. Liu, and Y. Xu, *J. Power Sources* **267**, 469 (2014)
[15] M. Winter, J.O. Besenhard, M.E. Spahr, and P. Novák, *Adv. Mater.* **10**, 725 (1998)
[16] C.K. Chan, H. Peng, G. Liu, K. McIlwrath, X.F. Zhang, R.A. Huggins, and Y. Cui, *Nat. Nanotechnol.* **3**, 31 (2008)
[17] X.H. Liu, L. Zhong, S. Huang, S.X. Mao, T. Zhu, and J.Y. Huang, *ACS Nano* **6**, 1522 (2012)
[18] J.K. Lee, K.B. Smith, C.M. Hayner, and H.H. Kung, *Chem. Commun.*, **46**, 2025 (2010)
[19] Y. Ma, R. Younesi, R.J. Pan, C.J. Liu, J.F. Zhu, B.Q. Wei, K. Edström, *Adv. Funct. Mater.* **26**, 6797 (2016)
[20] E. Greco et al., *J. Mater. Chem. A* **5**, 19306 (2017)
[21] S. Palumbo et al., *ACS Appl. Energy Mater.* (2019)

[22] J. Luo, X. Zaho, J. Wu, H.D. Jang, H.H. Kung, and J. Huang, *J. Phys. Chem. Lett.* **3**, 1824 (2012)

[23] Y. Ma, H.D. Asfaw, C.J. Liu, B.Q. Wei, K. Edström, *Nano Energy* **30**, 745 (2016)

[24] F. Jeschull, F. Lindgren, M. Lacey, F. Björefors, K. Edström, D. Brandell, *Journal of Power Sources,* **325** 513 (2016)

[25] A.E. del Rio Castillo et al., *Chem. Mater.* **30**, 506 (2018)

[26] C.-M. Park, and H.J. Sohn, *Adv. Mater.* **19**, 2465 (2007)

[27] F. Bonaccorso et al., *Science* **347**, 1246501 (2015)

[28] E. Quesnel et al., *2D Mater.* **2**, 030204 (2015)

[29] C. Zhang et al., *Nature Communications* **10**, 849 (2019)

[30] E.C. Evarts et al., *Nature* **526**, 93 (2015)

[31] B. Liu, Ji-G. Zhang, and W Xu, *Joule* **2**, 833 (2018)

[32] H. Sun, and K. Zhao, *J. Phys. Chem. C* **121**, 6002 (2017)

[33] J.M. Tarascon, and M. Armand, *Nature* **414**, 359 (2001)

[34] B. Kang, and G. Ceder, *Nature* **458**, 190 (2009)

[35] B. Scrosati, J. Hassoun, and Y.-K. Sun, *Energy Environ. Sci.* **4**, 3287 (2011).

[36] P. Rozier, and J.M. Tarascon, *J. Electrochem. Soc.* **162**, A2490, (2015)

[37] F.A. Susai, et al., *Advanced Materials*, **30** 1801348 (2018)

[38] X. Li, et al., *Chemistry of Materials* **30**, 2566 (2018)

[39] M. Ghorbanzadeh, et al., *Journal of Solid State Electrochemistry* **22**, 1155 (2018)

[40] J. Zheng, et al., *Advanced Energy Materials* **7**, 1601284 (2017)

[41] M. Gu et al., ACS Nano, **7**, 760 (2013)

[42] A. Mauger, M. Armand, C.M. Julien, K. Zaghib, *J. Power Sources*, **353,** 333 (2017)

[43] F. Zheng, M. Kotobuki, S. Song, M.O. Lai, and L. Lu, *J. Power Sources* **389,** 198 (2018)

[44] W. Zhou, Z. Wang, Y. Pu, Y. Li, S. Xin, X. Li, J. Chen, and J.B. Goodenough, *Adv. Mater.* **31** 1805574 (2018)

[45] Z. Gao, H. Sun, L. Fu, F. Ye, Y. Zhang, W. Luo, and Y. Huang, *Adv. Mater.* **30,** 1705702 (2018)

[46] C. Sun, J. Liu, Y. Gong, D.P. Wilkinson, and J. Zhang, *Nano Energy* **33**, 363 (2017)

[47] Q. Zhang, K. Liu, F. Ding, and X. Liu, *Nano Research* **10**, 4139 (2017)

[48] J. Mindemark, M.J. Lacey, T. Bowden, and D. Brandell, *Prog. Polym. Sci.* **81**, 114 (2018)

[49] H. Zhang, C. Li, M. Piszcz, E. Coya, T. Rojo, L.M. Rodriguez-Martinez, M. Armand, and Z. Zhou, *Chem. Soc. Rev.* **46** 797 (2017)

[50] J. Zhang, N. Zhao, M. Zhang, Y. Li, P. K. Chu, X. Guo, Z. Di, X. Wang, and H. Li, *Nano Energy* **28**, 447 (2016)

[51] Q. Wang, et al., *J. Electrochem. Soc.* **162**, 474 (2015)

[52] S. Waluś, et al., *Adv. Energy Mater.* **5**, 1500165 (2015)

[53] S.S. Zhang, *J. Power Sources* **231**, 153 (2013)

[54] H. Kim, et al., *Adv. Energy Mater.* **5**, 1501306 (2015)

[55] M.R. Busche, et al., *J. Power Sources* **259**, 289 (2014)

[56] Y.V. Mikhaylik, and J.R. Akridge, *J. Electrochem. Soc.* **151**, 1969 (2004)

[57] J.T. Lee, Y. Zhao, H. Kim, W.I. Cho, and G. Yushin, *J. Power Sources* **248**, 752 (2014)

[58] N. Ding, S.W. Chien, T.S.A. Hor, Z. Liu, Y. Zong, *J. Power Sources* **269**, 111 (2014)

[59] C. Jia-Jia, et al., *Electrochim. Acta* **55**, 8062 (2010)

[60] L. Carbone, et al., *ChemElectroChem* **4**, 209 (2017)

[61] B. Wang, B.; et al., *Chem. - A Eur. J.* **20**, 5224 (2014)

[62] S. Zhang, N. Li, H. Lu, J. Zheng, R. Zang, and J. Cao, *RSC Adv* **5** 509083 (2015)

[63] Q. Zhang, Y. Wang, Z.W. Seh, Z. Fu, R. Zhang, and Y. Cui, *Nano Lett.* **15**, 3780 (2015)

[64] R. Chen, T. Zhao, J. Lu, F. Wu, L. Li, J. Chen, G. Tan, Y. Ye, and K. Amine, *Nano Lett.* **13** 4642 (2013)

[65] J.L.Gómez-Urbano, J.L. Gomez-Camer, C. Botas, T. Rojo, D. Carriazo, *J. Power Sources* **412**, 408 (2019)

[66] J. Song, M.L. Gordin, T. Xu, S. Chen, Z. Yu, H. Sohn, J. Lu, Y. Ren, Y. Duan, and D. Wang, *Angew. Chemie - Int. Ed.* **54**, 4325 (2015)

[67] C. Barchasz, J.C. Leprêtre, F. Alloin, and S. Patoux, *J. Power Sources* **199**, 322 (2012)

[68] L. Carbone et al., submitted

[69] L. Grande, E. Paillard, J. Hassoun, J.B. Park, Y.J. Lee, Y.K. Sun, S. Passerini, and B. Scrosati, *Advanced Materials* **27**, 784 (2015)

[70] P. Stevens, G.Toussaint, G.Caillon, P. Viaud, P.Vinatier, C. Cantau Odile Fichet, C. Sarrazin and M. Mallouki *ECS Trans.* **28**, 1 (2010)

[71] C.J. Allen, J. Hwang, R. Kautz, S. Mukerjee, E.J. Plichta, M.A. Hendrickson, and K. M. Abraham, *J. Phys. Chem. C* **116**, 20755 (2012)

[72] Z. Ma, X. Yuan, L. Li, Z.-F. Ma, D.P. Wilkinson, L. Zhang, and J. Zhang, *Energy Environ. Sci.* **8**, 2144 (2015)

[73] Z. Ma, X.X. Yuan, L. Li, Z.F. Ma, D.P. Wilkinson, L. Zhang, J.J. Zhang *Energy & Environmental Science* **8**, 2144 (2015)

[74] S. Vankova, C. Francia, J. Amici, J. Zeng, S. Bodoardo, N. Penazzi, G. Collins, H. Geaney, and C. O'Dwyer, *ChemSusChem* **10**, 575 (2017)

[75] F. Wang, X. Li, *ACS Omega* **3**, 6006 (2018)

[76] S. Martinez Crespiera, D. Amantia, E. Knipping, C. Aucher, L. Aubouy, J. Amici, J. Zeng, U. Zubair, C. Francia, and S. Bodoardo, *Journal of Applied Electrochemistry* **47**, 497 (2017); Y.-G. Huang, J. Chen, X.-H. Zhang, Y.-H. Zan, X.-M. Wu, Z.-Q. He, H.-Q. Wang, Q.-Y. Li, *Chemical Engineering Journal* **296**, 28 (2016); K. Wang, S. Xu, Q. Zhu, F. Du, X. Li, X. Wei, J. Chen, *Dalton Trans.*, **44**, 8678 (2015); H. Wang, H. Chen, H. Wang, L. Wu, Q. Wu, Z. Luo, F. Wang, *Journal of Alloys and Compounds*, **780**, 107 (2019)

[77] J. Amici, C. Francia, J. Zeng, S. Bodoardo, and N. Penazzi, *J. Appl. Electrochem*. **46**, 617 (2016)

[78] N. Ortiz-Vitoriano, N. E. Drewett, E. Gonzalo, and T. Rojo, *Energy Environ. Sci.*, **10**, 1051 (2017)

[79] J. Xiao, D. Mei, X. Li, W. Xu, D. Wang, G. L. Graff, W. D. Bennett, Z. Nie, L. V. Saraf, I. A. Aksay, J. Liu and J. G. Zhang, *Nano Lett.*, **11**, 5071 (2011)

[80] M. Enterria, C. Botas, J.L. Gomez-Urbano, B. Acebedo, J. M. Lopez del Amo, D. Carriazo, T. Rojo, and N. Ortiz-Vitoriano *J. Mater. Chem.* A **6**, 20778 (2018)

[81] F.A. Faber, A. Lindmaa, O. A. von Lilienfeld, and R. Armiento *Phys. Rev. Lett.* **117**, 135502 (2016)

[82] N Mounet, et al., *Nature Nanotechnology* **13**, 246 (2018)

[83] S. Das Sarma, D.-L. Deng, L.-M- Duan *Physics Today* **72**, 48 (2019)